# CLASSICAL MECHANICS WITHOUT ABSOLUTE SPACE

D. Lynden-Bell and J. Katz*

Institute of Astronomy, The Observatories, Cambridge, U.K.
*Racah Institute of Physics, Hebrew University, Jerusalem, Israel

**Abstract** A relative mechanics with no absolute space is shown to be equivalent to Newtonian mechanics applied in a universe of zero net angular momentum.

Closed spaces in General Relativity have no angular momentum and shrivel to one point as the mass-energy contained tends to zero, so obeying Mach's principle on the origin of inertia.

Introduction

Newton [1] claimed that rotation was absolute and invented the concept of Absolute Space independent of the motion of the bodies therein. Mach [2] pointed out that Newton's experiments in no way excluded the idea that dynamical rotation was relative to distant masses in the Universe. Here we outline a classical theory in which only relative motions are meaningful, (no absolute space, no absolute rotation) which agrees precisely with Newtonian mechanics when the latter is applied in a world in which the whole Universe has no net angular momentum. Our *mechanics* is independent of frame rotation. The Universe breaks that symmetry.

In General Relativity there is a corresponding result that applies to closed universes with hyperspherical topology. In those the vanishing of the angular momentum is an inevitable consequence of closure.

Background Observations

Barrow et al. [3] have refined the strict limit on the rotation of the Universe first derived by Hawking [4] and collaborators [4,5] from the isotropy of the microwave background radiation. They deduce that the Universe has turned by less than $10^{-12}$ of one rotation in the time since the Big Bang. This amazingly stringent limit rests on the assumption that the microwave background radiation was last scattered at a redshift around $z = 1000$. The directly observed local inertial frame defined from the celestial mechanics of the solar system (essentially using it as a giant gyroscope) agrees with the frame defined by the directions of the extra-galactic nebulae. The accuracy is better than 3 rotations since the Big

Bang. Thus there is excellent evidence both indirect and direct that the Universe does not rotate as a whole. We infer that the total angular momentum of the Universe is zero.

Lagrangian for Newtonian and Relative Mechanics

Newtonian mechanics can be derived concisely from the action principle $\delta \int L_N dt = 0$, where the Lagrangian $L_N = T_N - V$. The Newtonian kinetic energy, $T_N = \sum \frac{1}{2} m_i \mathbf{v}_i^2$, depends on the reference frame while the potential energy $V$ does not since it involves only the relative positions. Furthermore, $T_N$ is a single sum over particles, while $V$ is a double or multiple sum over pairs as in the gravitational potential energy $V_g = -\sum_{i<j} \sum G m_i m_j / |\mathbf{r}_i - \mathbf{r}_j|$. Relative to the centre-of-mass frame, the kinetic energy $T = T_N - \frac{1}{2} M \mathbf{u}^2 = \sum \frac{1}{2} m_i (\mathbf{v}_i - \mathbf{u})^2$, can be rewritten in the form $T = \sum_{i<j} \sum \frac{1}{2} (m_i m_j / M)(\mathbf{v}_i - \mathbf{v}_j)^2$ which is more similar to $V_g$. We have written $M = \sum m_i$ for the total mass and $\mathbf{u} = M^{-1} \sum m_i \mathbf{v}_i$ for the velocity of the barycentre. Hereafter $m_i m_j / M = m_{ij}$. $T_N$ is invariant under the time-independent displacement $\mathbf{r} \to \mathbf{r} + \boldsymbol{\Delta}$ but $T$ is invariant under the far larger group of arbitrary time-dependent displacements $\mathbf{r} \to \mathbf{r} + \boldsymbol{\Delta}(t)$. If $L = T - V$ is used as Lagrangian in place of $L_N$, then it gives no equation of motion for the barycentre of the whole Universe, so we may choose it to move in any way we like! If we *choose* it to move uniformly, then we recover precisely Newton's equations so, apart from our new freedom to choose the movements of the barycentre of the Universe, the dynamics based on $L$ in place of $L_N$ is equivalent to Newton's.

While $T$ involves only what we commonly call the relative velocities of the bodies, nevertheless it does depend on the rotation of the reference frame. Denoting time derivatives relative to a frame Newtonianly rotating with angular velocity $\boldsymbol{\Omega}(t)$ by a dot, we have $\mathbf{v}_i - \mathbf{v}_j = \dot{\mathbf{r}}_i - \dot{\mathbf{r}}_j + \boldsymbol{\Omega} \times (\mathbf{r}_i - \mathbf{r}_j)$.

The kinetic energy perceived relative to such a frame is

$$T^* = \sum_{i<} \sum_j \tfrac{1}{2} m_{ij} \left(\dot{\mathbf{r}}_i - \dot{\mathbf{r}}_j\right)^2 = \sum_{i<} \sum_j \tfrac{1}{2} m_{ij} \left[\mathbf{v}_i - \mathbf{v}_j - \boldsymbol{\Omega} \times (\mathbf{r}_i - \mathbf{r}_j)\right]^2 . \tag{1}$$

The least value that $T^*$ takes in any frame is found by minimising it over all choices of $\boldsymbol{\Omega}(t)$. The minimising $\boldsymbol{\Omega}$ is readily found to obey

$$\underline{\underline{I}} \cdot \boldsymbol{\Omega} = \mathbf{J} = \sum_{i<} \sum_j m_{ij} \left(\mathbf{r}_i - \mathbf{r}_j\right) \times (\mathbf{v}_i - \mathbf{v}_j) \tag{2}$$

where $\underline{\underline{I}}$ is the time dependent moment of inertia about the barycentre about which the angular momentum is $\mathbf{J}$. In barycentric coordinates $\mathbf{J} = \sum m_i \mathbf{r}_i \times \mathbf{v}_i$ and $\underline{\underline{I}} = \sum m_i \left( r_i^2 \underline{\underline{\delta}} - \mathbf{r}_i \mathbf{r}_i \right) = \sum_{i<j} \sum m_{ij}[(\mathbf{r}_i - \mathbf{r}_j)^2 \underline{\underline{\delta}} - (\mathbf{r}_i - \mathbf{r}_j)(\mathbf{r}_i - \mathbf{r}_j)]$. Here $\underline{\underline{\delta}}$ is Kronecker's unit tensor with components $\delta_{ab}$. Many double summations occur hereafter, so we adopt a more convenient notation for sums over all pairs of particles. We denote by $\alpha$ a pair $ij$ and let $\alpha$ run from 1 to $\frac{1}{2}N(N-1)$, where $N$ is the number of particles; we write $\mu_\alpha = m_{ij} = m_i m_j / M$ and put $\mathbf{r}_\alpha = \mathbf{r}_i - \mathbf{r}_j$. In this notation the last expression for $\underline{\underline{I}}$ becomes $\underline{\underline{I}} = \sum_\alpha \underline{\underline{I}}_\alpha$ where $\underline{\underline{I}}_\alpha = \mu_\alpha \left( r_\alpha^2 \underline{\underline{\delta}} - \mathbf{r}_\alpha \mathbf{r}_\alpha \right)$. The minimised perceived kinetic energy is

$$T_m^* = T - \tfrac{1}{2} \mathbf{\Omega} \cdot \underline{\underline{I}} \cdot \mathbf{\Omega} = T - \tfrac{1}{2} \mathbf{J} \cdot \underline{\underline{I}}^{-1} \cdot \mathbf{J} = \sum_\alpha \mu_\alpha \left\{ \dot{r}_\alpha^2 + [(\boldsymbol{\omega}_\alpha - \mathbf{\Omega}) \times \mathbf{r}_\alpha]^2 \right\} \quad (3)$$

where $\mathbf{v}_\alpha = d\mathbf{r}_\alpha / dt = \dot{r}_\alpha \hat{\mathbf{r}}_\alpha + \boldsymbol{\omega}_\alpha \times \mathbf{r}_\alpha$, so $\boldsymbol{\omega}_\alpha$ is the angular velocity of the line joining the pair labelled $\alpha$ and $\hat{\mathbf{r}}_\alpha$ is the unit vector along $\mathbf{r}_\alpha$. From (2) $\underline{\underline{I}} \cdot \mathbf{\Omega} = \mathbf{J} = \sum \underline{\underline{I}}_\beta \cdot \boldsymbol{\omega}_\beta$ so $\mathbf{\Omega} = \underline{\underline{I}}^{-1} \cdot \sum \underline{\underline{I}}_\beta \cdot \boldsymbol{\omega}_\beta$ and (3) may be rewritten

$$T_m^* = \sum_\alpha \mu_\alpha \left[ \dot{r}_\alpha^2 + \{ [\underline{\underline{I}}^{-1} \cdot \sum_\beta \underline{\underline{I}}_\beta \cdot (\boldsymbol{\omega}_\alpha - \boldsymbol{\omega}_\beta)] \times \mathbf{r}_\alpha \}^2 \right]. \quad (4)$$

In this final form $T_m^*$ is explicitly independent of the frame used to calculate it because the non-vectorial $\dot{r}_\alpha$ is just the rate of change of a distance, while on a change of frame rotation rate both $\boldsymbol{\omega}_\alpha$ and $\boldsymbol{\omega}_\beta$ change in the same way leaving $\boldsymbol{\omega}_\alpha - \boldsymbol{\omega}_\beta$ invariant.

We use $L^* = T_m^* - V$ as the Lagrangian of our new formulation of dynamics [6]; it is invariant under all time-dependent changes of the orthogonal reference frame so its invariance group is far greater than that of the Newtonian $L_N$. When $T_m^*$ is subjected to a small variation, one of the two $\mathbf{J}$ in expression (3) will be left unvaried in each term by which $\delta T_m^*$ differs from $\delta T$. Hence when used of a universe in which $\mathbf{J}$ is zero the equations derived from $L^*$ will be the same as those derived from $L$. We showed earlier that those were equivalent to Newton's when we chose the barycentre of the universe to move uniformly. Thus for our Universe which evidently has no Newtonian angular momentum, the relative Lagrangian $L^*$ correctly describes Classical dynamics.

Since $L^*$ only involves relative orientations, separations, and their rates of change absolute space no longer appears in this new formulation of dynamics. This is therefore the solution to the classical problem.

Comments

Why is it then that we can tell from the dynamics of a subsystem whether our frame is rotating or not? The answer lies in our expression for $T_m^*$ which is not a simple sum over independent subsytems; however, if we choose a frame such that $\sum \underline{\underline{I}}_\alpha \cdot \boldsymbol{\omega}_\alpha = 0$ and $\sum m_i \dot{\mathbf{r}}_i =$ constant, then $T_m^*$ becomes the simple sum $\sum \frac{1}{2} m_i \dot{\mathbf{r}}_i^2$, so the dynamics of the subsytem in such a frame is, in the absence of large scale gravity gradients, independent of the rest of the Universe. A different way of saying this is that the rest of the Universe only intrudes on local dynamics through the inertial frame defined by the motions of the Universe. *The dynamics is frame-independent and purely relative, but the Universe is not.*

Other interesting discussions of quasi-classical theories are due to Zanstra [7], Schrödinger [8], and particularly Barbour & Bertotti [9,10], whose second paper briefly considers in §3 a theory that can be shown to be mathematically equivalent to that given here.

Relativistic Extension

Extension of these ideas to general relativity is a much harder problem which has been greatly elucidated in our more technical paper [11]. Essentially similar conclusions hold for any closed universe and all closed universes are shown to have no angular momentum. Mach [12], Einstein [13] and also Bondi [14] were much puzzled by the conundrum that inertia must be almost unchanged in an almost empty Minkowski space since light cones define unaccelerated axes [15]. This is contrary to their concept that "there is no inertia of mass against space but only inertia of mass against mass". If only closed solutions of Einstein's equations are admitted as physical (as only they obey Einstein's closure condition), then this conundrum is beautifully removed. At maximum extension any closed Friedmann Universe has $R = 2GMc^{-2}$ where $M$ is the mass-energy in the hemi-hypersphere of radius of curvature $R$. As closed Friedmann universes with less and less mass-energy are considered, the maximum radius becomes smaller, the mean density gets larger and the time between the Big Bang and the Big Crunch diminishes. Finally, as all mass-energy is removed, space-time shrivels to a point leaving no space, no time, and no inertia. Closed universes with density greater than the critical one behave according to Mach's Principle.

But consider the similar experiment of comparing open universes with less and less mass. They lose their gravitational retardation earlier and earlier and become more and more like the empty Milne model, which is in Minkowski Space. In them the space-time determines the light cones and all the inertia when all the mass is removed. The Open universes do not obey Mach's Principle. Thus closed boundary conditions are necessary to make Einstein's equations Machian. Empty Minkowski space never arises as it disobeys the closure condition.